\input{aipcheck}

\documentclass[
,draft            
]
  {aipproc}
\usepackage{epsfig}
\usepackage{graphicx}
\usepackage{bm}

\newcommand{\be}{\begin{equation}}

\newcommand{\ee}{\end{equation}}
\newcommand{\bea}{\begin{eqnarray}}
\newcommand{\eea}{\end{eqnarray}}

\def\qq{Q\!\!\!\!\!Q}

\newcommand{\nn}{\nonumber}
\newcommand{\de}{\partial}
\layoutstyle{6x9}

\begin{document}
\title{QCD, hadrons and beyond}


\author{G. Nardulli}{
  address={Department of Physics, University of Bari and
INFN-Bari, Italy} }

\begin{abstract}
I give a summary of Section E of the sixth edition of the
Conference {\it Quark confinement and the hadron spectrum}. Papers
were presented on different subjects, from spectroscopy, including
pentaquarks and hadron structure, to new physics effects (non
commutative field theories, supersymmetry and extra dimensions)
and the problem of color confinement, both in ordinary Yang-Mills
models and in supersymmetric Yang-Mills.
\end{abstract}

\maketitle

\section{Pentaquarks, Spectroscopy and  Hadron structure}

Starting from January 2003 several experiments have produced
evidence for a few baryonic resonances whose simplest
interpretation in terms of the quark model is that of a bound
state of five quarks, more exactly four quarks and an antiquark.
The first observed pentaquark has been the baryonic resonance
$\Theta^+(1540)$, reported by several experiments: LEPS
\cite{Nakano:2003qx}, DIANA \cite{Barmin:2003vv}, CLAS
\cite{Stepanyan:2003qr,Kubarovsky:2003fi}, SAPHIR
\cite{Barth:2003es}, HERMES \cite{Airapetian:2003ri}, as
 well as by analyses of old bubble chamber experiments \cite{Asratyan:2003cb}.
 Several new experimental results on this state have been presented at this conference,
 and a lively discussion took place also in Section E. All the experiments
giving evidence of $\Theta^+(1540)$ show that this resonance
decays into $K^+n$ or $K^0_sp$ with a width compatible with
experimental resolution ($\Gamma\sim$ a few MeV). The former
decay, with strangeness $S=+1$ and baryonic number $B=1$  is
exotic. In principle one might model this state as a molecule
$K^+n$; this interpretation would require an interplay of an
attractive interaction with a range of $\sim 1$ fermi and the
centrifugal barrier ($\ell\neq 0$); however such a model is
disfavored since it would produce too large widths. As remarked
above, the simplest quark model interpretation is that of a
pentaquark, i.e. an exotic state formed by five quarks: $udud\bar
s$. Other narrow exotic cascade states, e.g. a $\Xi^{--}$ state
with quantum numbers $B=1,Q=S=-2$, and also a $\Xi^{-}$ and
$\Xi^{0}$ state have been reported by the NA49 Collaboration, see
\cite{Alt:2003vb}. Also these signals can be interpreted as
pentaquark states, e.g. for $\Xi^{--}$, $dsds\bar u$.

In the first year there has been an almost continuous flow of
experimental results, but, starting from January 2004, new data
appeared, many with negative results \cite{Pochodzalla:2004up},
e.g. BES, OPAL, PHENIX, DELPHI, ALEPH, CDF, BaBar, E690.

Negative results from the Hera-B experiment were presented in
Section E by M.Medinnis. The search for $\Theta^+(1540)$ and
$\Xi^{--}$  in $pA$ collisions at 920 GeV from this experiment
only resulted in upper bounds. More precisely, looking at the
$pK^0_s$ invariant mass, Hera-B finds the upper limit \be {\cal
B}d\sigma/dy|_{y=0}= 3.7\mu b/N\ee at 1530 MeV/$c^2$
 and\be {\cal B}d\sigma/dy|_{y=0}= 2.4\mu b/N \ee at 1540 MeV/$c^2$. The upper limit for $\Xi^{--}$
 production is  $2.5\mu$b/N at 1862 MeV/$c^2$, i.e. in the mass region where
 NA49 finds positive results.

 J. Engelfried reported results on the
 search of strange and charmed pentaquark states at
 Selex. This collaboration finds no evidence of  the strange pentaquark
 $\Theta^+$, while for the charmed pentaquark $\Theta_c$ no
 conclusion can be drawn yet.

 The charmed pentaquark state $\Theta^0_c$
has been searched by the H1 and ZEUS Collaborations at DESY.
 L. Gladilin reported results
 from these two experiments. H1 \cite{Aktas:2004qf} finds a 5$\sigma$ signal
 in deep inelastic scattering and a signal in photoproduction
 at the same mass ($\sim 3.1$ GeV). This narrow anticharmed state is seen through its decays
 into $D^{*-} p + c.c.$ and its minimal quark content is $udud\bar c$. On
 other hand  ZEUS does not find it \cite{Chekanov:2004qm}. As to the
 strange
 pentaquark $\Theta^+$,
 ZEUS \cite{Chekanov:2004kn} observes $221\pm 48$
  events in the channel $pK^0_s$ with a mass of the
 $\Theta^+$ equal to $ 1521.5\pm 1.5$ MeV. On the other hand no signals of the
 $\Xi^{--}$ pentaquark is found by this collaboration.

Much experimental effort is expected in the near future to clarify
these experimental issues. The origin of the discrepancies might
be in the difference of production mechanisms, leading to
different yields in different experiments. A careful analysis of
the different assumptions in the experimental analyses would be
certainly welcome, in particular those related to the  kinematical
cuts. In any event high statistics experiments should provide an
answer in the near future. For the time being we can certainly
assert that the appearance of exotic states, coming after years of
fruitless experimental researches of exotica, has revived
theoretical interest in QCD spectroscopy and its low energy
models. Pentaquark states were indeed predicted long ago in the
framework of the Chiral Soliton Model
\cite{Manohar:1984ys,Chemtob:1985ar}, which is an extension to
three flavors \cite{Witten:1983tx,Adkins:1983ya,Guadagnini:1984uv}
of the Skyrme model \cite{Skyrme:1961vq,Skyrme:1962vh}. In the
Chiral Quark Soliton Model \cite{Walliser:1992vx,Diakonov:1997mm}
all baryonic states are interpreted as arising from quantizing the
chiral nucleon soliton and the pentaquark emerges as the third
rotational excitation with states belonging to an antidecuplet
with spin $s=1/2$. Other interpretations have been proposed after
the first results on $\Theta^+(1540)$, most notably the one of
Jaffe and Wilczek \cite{Jaffe:2003sg,Jaffe:2004zg} who propose
that the $\Theta^+$ comprises two highly correlated $ud$ pairs
(diquarks: $\qq$) and an $\bar s$. Diquarks properties are similar
to those of the diquark condensates of QCD in the high density
color-flavor-locking (CFL) phase \cite{Alford:1998mk}. Both
diquarks are in spin 0 state, antisymmetric in color and flavor.
Together they produce a $\qq\qq$ state in the flavor-symmetric
$\bf 6_f$ that must be antisymmetric in color and in $p-$wave to
satisfy Bose statistics. When combined with the antiquark the
diquarks produce a $\bf \overline{10}_f$ with spin 1/2 and
positive parity (they can also produce a $\bf 8_f$, and mixing is
possible).

The hypothesis that the attractive interaction in the
antisymmetric color channel may play a role both at low and high
density quark matter is especially interesting in the light of the
quark hadron continuity which has been suggested
\cite{Schafer:1998ef} to exist between the CFL and the
hypernuclear phase.
 Due to the formation of the CFL condensate that breaks color,
 flavor and the electric charge, though preserving
a combination of the electric charge and of the color generator
$T_8$, the physical states are obtained by dressing the quarks by
diquarks. The result is that in this phase eight quarks have
exactly the same quantum numbers of baryons. Also the ninth quark
corresponds to a singlet with a gap which is twice the gap of the
octet. The same phenomenon takes place for the other states. For
instance, the gluons are dressed by a pair $\overline{\qq}\qq$
giving rise to vector states with the same quantum number of the
octet of vector resonances ($\rho$, etc.).

\begin{figure}[h] \centerline{
\epsfxsize=12cm\epsfbox{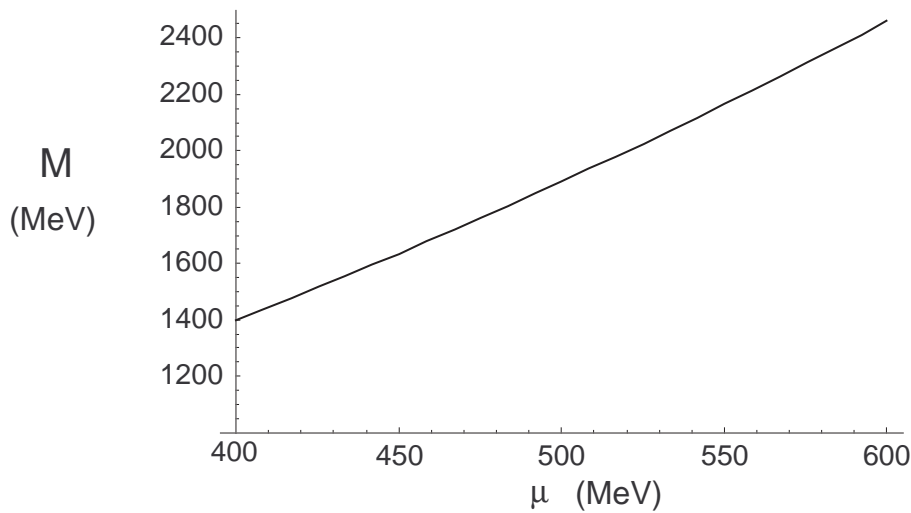} } \caption { The soliton
mass $M$ at finite density in the CFL phase as a function of the
baryonic chemical potential $\mu$, for the value of the gap
$\Delta=40 $ MeV.\label{fig:1} }
\end{figure}

Quark-hadron-continuity plays a role in relating quark and baryons
in the low-lying octet. Apparently it also matters in assigning a
role to diquark attraction  at zero baryonic densities. In a
recent paper \cite{Casalbuoni:2004xs} it has been suggested  that
another sign of it is the existence of baryon chiral solitons also
at finite density. The mechanism of its formation is based on the
existence of condensates giving rise to color superconductivity
and Nambu Goldstone Bosons (NGB) associated to the breaking of
global symmetries. One starts with the effective lagrangian for
the Nambu-Goldstone bosons written in terms of the fields bosonic
fields $X$ and $Y$ associated to the right handed and left handed
spin 0 condensates \cite{Casalbuoni:2004xs}: \bea {\cal
L}&=&-\frac{F_T^2}4{\rm Tr}\left[\left( X\de_0 X^\dagger- Y\de_0
Y^\dagger)^2\right)\right]-\alpha_T\frac{F_T^2}4{\rm
Tr}\left[\left( X\de_0  X^\dagger+ Y\de_0
Y^\dagger+2 g_0)^2\right)\right]\nn\\
&&+\frac{F_S^2}4{\rm Tr}\left[\left|  X{\bm\nabla} X^\dagger-
Y{\bm\nabla} Y^\dagger\right|^2\right]+\alpha_S\frac{F_S^2}4{\rm
Tr}\left[\left| X{\bm\nabla} X^\dagger+ Y{\bm\nabla}\hat
Y^\dagger+2 {\bf g}\right|^2\right] \nn\\&&+\frac 12
(\de_0\phi)^2-\frac{v_\phi^2}2|{\bm\nabla}\phi|^2-\frac
1{g_s^2}Tr[{\epsilon \bf E}^2-\frac 1{\lambda}{\bf
B}^2].\label{lagrangian1}
 \eea where the gluon strength is
 \be
 F_{\mu\nu}=\de_\mu g_\nu-\de_\nu g_\mu-[g_\mu,g_\nu]\ee
 and
 \be E_i=F_{0i},~~~~B_i=\frac 12\epsilon_{ijk}F_{jk}\ .\ee The parameters
 $\epsilon$ and $\lambda$ are the dielectric constant and the
 magnetic permeability of the dense condensed medium.
In the CFL vacuum the gluons $g_0^a$ and $g_i^a$ acquire Debye and
Meissner masses given by \be m_D^2=\alpha_Tg_s^2
F_T^2,~~~m_M^2=\alpha_Sg_s^2 F_S^2=\alpha_Sg_s^2v^2 F_T^2.
\label{masses}\ee where \be v^2=\frac{F_S^2}{F_T^2}\ .\ee It
should be stressed that these are not the true rest masses of the
gluons, since there is a large wave function renormalization
effect making the gluon masses of the order of the gap $\Delta$,
rather than $\mu$ \cite{Casalbuoni:2000na}. One  can decouple the
gluons solving their classical equations of motion neglecting the
kinetic term. The result from Eq. (\ref{lagrangian1}) is \be
g_\mu=-\frac 12 \left( X\de_\mu X^\dagger+ Y\de_\mu
Y^\dagger\right). \label{6.18}\ee By substituting this expression
in Eq. (\ref{lagrangian1}), and performing a gauge rotation to get
$Y=1$, one gets \be {\cal L}=\frac{F_T^2}4\left({\rm
Tr}[\dot\Sigma\dot\Sigma^\dagger]-v^2{\rm
Tr}[\vec\nabla\Sigma\cdot\vec\nabla\Sigma^\dagger]\right)+ \frac
12\left(\dot\phi^2-v_\phi^2|\vec\nabla\phi|^2\right)-\frac
1{g_s^2}Tr[{\epsilon \bf E}^2-\frac 1{\lambda}{\bf B}^2].
\label{6.17}\ee with \be E_i=\frac 1
4[\Sigma\de_0\Sigma^\dagger,\Sigma\de_i\Sigma^\dagger],~~~~~B_i=\frac
1
8\epsilon_{ijk}[\Sigma\de_j\Sigma^\dagger,\Sigma\de_k\Sigma^\dagger].\ee
Apart  for the breaking of the Lorentz symmetry, one recognizes in
the first term the chiral lagrangian and, in the last one,  the
Skyrme term \cite{Skyrme:1962vh}. This effective lagrangian
enforces the idea of the quark-hadron continuity between the CFL
and the hypernuclear matter phase with three flavors. A numerical
estimate of the soliton mass based on these assumptions is in Fig.
\ref{fig:1} (for  $\Delta=40 $ MeV). \vskip1cm
Around $400 $ MeV the soliton mass is in the range of
1200-1400 MeV, which, in the light of the quark-hadron-continuity,
is in the right ball-park. If pentaquarks exist at zero density,
they should be connected  continuously with states comprising two
diquarks and an antiquark existing at finite density.

Spectroscopy and the study of the hadron structure may appear as
old-fashioned sectors of hadron physics. They are however from
time to time also source of surprises in theoretical physics. We
have discussed the recent excitement about pentaquarks. Another
much debated subject a decade ago concerned  the proton spin. On
this subject U.  D'Alesio gave a talk on the role of the intrinsic
partonic $k_T$. The dependence on intrinsic partonic $k_T$   in
parton distribution function (pdf)  can be parameterized by a
gaussian shape and is born by a twist 2 operator (Sivers) whose
effect on asymmetries is competitive with another source of
asymmetry, the so called Collins effect. Its presence in the pdf
is of interest in single spin asymmetries for inclusive production
by high energy transversely polarized hadron hadron scattering.
D'Alesio and collaborators \cite{dalesio1,dalesio2}  give
predictions for several transverse single spin asymmetries, e.g.
at RHIC. In particular  single spin  asymmetries in Drell Yan
processes  can provide a tool to extract quark Sivers distribution
functions, while the inclusive production of $D$ by high energy
transversely polarized hadron hadron scattering could be a  tool
to extract gluon distribution function. Related to this study is
another talk given in Section E, by I. Vukotic for the HERA-B
Collaboration, who presented results for charmonium and open charm
production obtained by this collaboration. In particular a new
limit \be \sigma(D^0\to\mu^+\mu^-)= 2.0 \times10^{-6}\hskip1cm
(90\% CL)\ee was presented, currently  the best published upper
limit for this decay.
\section{New physics effects}
The discussion on soliton states at finite density in the CFL is a
useful reminder that actually we  do not yet know the true ground
state of QCD at intermediate densities, i.e. those corresponding
to quark chemical potentials $\mu\sim 400$ MeV. It is possible
that the QCD  ground state  is characterized, at these densities,
by inhomogeneous color superconductivity. This state is called
LOFF state \cite{LO}, \cite{FF}, \cite{Casalbuoni:2003wh}. Its
prominent feature is diquark condensation with non vanishing total
momentum of the Cooper pair, so that  $\Delta=\Delta({\bf r})$.
The dynamics of these phases was investigated by P. Castorina in
the framework of non-commutative field theories (NCFT) with
cut-off \cite{Castorina:2004xb,Castorina2}.

The hypothesis that space coordinates do not commute:\be
[x_\mu,x_\nu]=\Theta_{\mu\nu}\label{sn}\ ,\ee
    can be traced back to Snyder \cite{Snyder:1946qz}.
   The condition (\ref{sn}) can be  realized in quantized motions of particles
   in strong
    magnetic field $H$ and the non commuting coordinates can be
    appropriately chosen
    on a  plane  perpendicular   to  $
    H$.
Castorina gave a talk on this subject with on overview on NCFT.
 Several cases of NCFT have been studied so far. They are based on
 the use of the Moyal product of fields:\be
 (f\star g)(x)= \exp\left\{\displaystyle i\frac{\Theta_{\alpha\beta}}2
 \,\partial_\alpha\partial^\prime_\beta\right\}\
 f(x)g(x^\prime)_{x=x^\prime}\ee
that implements non commutativity by means of the parameters
$\Theta_{\alpha\beta}$. Castorina and collaborators study
transitions from ordered phases with homogeneous order parameters
to phases with inhomogeneous order parameters. NCFT with cutoff
$\Lambda$ are used as an effective approach to describe the
dynamical mechanism underlying these transitions. They consider
two applications, one for $\lambda\phi^4$, in the context of
Bose-Einstein Condensation, and another one for Nambu Jona Lasinio
four-fermion coupling, that can be used either in the context of
superconductivity or for spontaneous  breaking of chiral symmetry.
In the latter case one considers: \be {\cal
L}=i\bar\psi\gamma\cdot\partial\psi+g\bar\psi_\alpha\star
\psi_\alpha\star\bar\psi_\beta\psi_\beta-g\bar\psi_\alpha\star\bar\psi_\beta
\psi_\alpha\star\psi_\beta\ .\ee For $g$ larger than a critical
value $g_c$ one has chiral symmetry breaking to a phase with
inhomogeneous order parameter. The space modulation of the gap is
similar to the LOFF case with a Cooper pair momentum $P\propto
1/\Theta\Lambda^2$.

C. Corian\`o and A. Feo gave talks on supersymmetric models. In
the analysis of spectra and hadron multiplicities for collisions
induced by Ultra High Energy Cosmic Rays (UHECR) it is important
to model the effects of new physics, to get a quantitative
understanding of their role in experimental observables. Corian\`o
presented results on the modifications induced by SUSY models not
only on UHECR but also in deep inelastic scattering. The
simulations he presented \cite{Cafarella:2004ws,Cafarella:2004hg}
include effects due to low energy gravity scales induced by extra
dimensions. Numerical effects can be significant indeed, and might
manifest themselves in new generation experiments, be they
collider physics experiments or cosmic rays observations.

Feo presented a study of dynamical  breaking of supersymmetry by
non perturbative lattice techniques, using the hamiltonian
formalism in a class of $d=2, N=1$ Wess Zumino models
\cite{feo1,feo2} (see also \cite{Feo:2002yi}).
 Their study includes an analysis of the phase
diagram by analytical strong coupling expansions and numerical
simulations. All results with cubic prepotential indicate unbroken
SUSY, while for quadratic potential \be V= \lambda_2\phi^2 +
\lambda_0\ee they  confirm the existence of two phases. At high
$\lambda_0$ there is a phase characterized by broken SUSY with
unbroken $Z_2$. At low $\lambda_0$ SUSY is unbroken SUSY and $Z_2$
is broken. The critical value of $\lambda_0$  they find is
\be\lambda_0^c=-0.48 \pm 0.01\ .\ee
\section{Confinement in Yang-Mills and Super Yang-Mills}

The last two talks I wish to summarize were presented in Section E
by K. Konishi and A. Niemi. They have in common the study of
topological effects in the discussion of confinement in Yang Mills
theories. As is well known, the simplest gauge model with monopole
solution is  the Georgi-Glashow model, based on the group
$G=SU(2)$ and containing a triplet of Higgs fields: \be
  {\cal      L} = -\frac 1 4{F}^2+\frac 1 2 (D \phi)^2+  \lambda(\phi^2-v^2)^2
  \ee where  \be  D_\mu\phi^a =\partial_\mu\phi^a + g\epsilon_{abc}
  A^b_\mu\phi^c\ .\ee

After symmetry breaking  a symmetry subgroup (e.m.) remains:
$H=U(1)_{e.m.}$.  The model has soliton solutions with charges \be
q=0\,,\hskip2cm g_M=\frac 1 g \neq 0\ .\ee

The model can be generalized to other groups $G,\,H$ usually in
the small $\lambda$ limit (Bogomol'nyi, Prasad, Sommerfield). It
is a simplified model but full of interest. Among the other
things, it teaches us that microscopic variables in the lagrangian
(e.g. in $\cal  L$ the fields $ \phi,\, A$) do not necessarily
coincide with observed quanta, that are massive gauge bosons and
monopoles. In QCD similar differences  arise between microscopic
(quarks, gluons) and macroscopic (hadrons) degree of freedoms.
Which ideas on confinement arise from this analogy? A popular
vision of confinement is based on the idea that two color charges
at large distances $R$ form a flux tube, i.e. a string like,
one-dimensional object with an energy $V (R) \sim \sigma R$.
Confinement might be explained by analogy with type II
superconductors. Let us consider two magnetic monopoles. Since
magnetic flux is conserved and  cannot vanish everywhere, it
remains confined to vortex lines (Abrikosov vortices)
characterized by \be E/R\simeq \sigma= const.\ee  Now in QCD one
needs chromoelectric, not chromomagnetic flux tubes, i.e. one
needs Cooper condensation of pairs of magnetic charges (dual
Meissner effect). In normal QCD magnetic monopoles do not exist as
particles, but in supersymmetric QCD (SQCD)  they can exist, as
first shown by Seiberg and Witten.

Konishi discussed several models \cite{kon,kon1} of monopole
confinement by vortices. After the construction of  nonabelian BPS
vortices he proves  that nonabelian monopoles  occur as infrared
degree of freedom (e.g. in $N=2, SU(N_c)$ SQCD). This suggests
that softly broken $N=2$ theories might  model QCD confinement.
The existence of nonabelian monopoles in these models is
essentially a quantum mechanical phenomenon. A peculiar feature of
these models is that massless flavor symmetry is important to keep
$ H$ unbroken and for being part of the dual transformation
itself.

Niemi discussed a model with glueballs as closed strings
\cite{niemi}. These topological solutions are stabilized by the
existence of twists or knots. Their stability  follows from
topological considerations, e.g. the existence of twist or knots
of the closed string. The model is based on the Yang-Mills theory
with two colors. In the infrared  regime one separates $A_\mu$ in
a $U(1)$ e.m. component $A_\mu$ and charged $A^\pm_\mu$. By these
components a composite gauge field $\Gamma_\mu$ can be constructed
for an internal group $U(1)_\Gamma$.
    The model can be casted in a form similar to a
    dual superconductor with 2 condensates
    (as in metallic hydrogen  or  high $T_c$ superconductors).
    The vortex lines have  a two-sheet structure which allows
    for twisting and knots. Stability of the closed strings
    follows from  this non trivial
topological structure.

These two talks have shown once again that topological solitons
(vortex lines) constitute  a lively subject of study. Therefore a
deeper understanding of the topological structure of QCD, or some
SUSY extension of QCD, can shed light on the dynamics of color
confinement, perhaps rendering this {\it millennium problem} (see
the site http://www.claymath.org/millennium) eventually solvable.

\begin{theacknowledgments}
  It is a pleasure to thank N. Brambilla, U. D'Alesio, G. Prosperi
and all the organizers for this beautiful conference in a
spectacular site.
\end{theacknowledgments}

\end{document}